\documentstyle[epsfig,preprint,aps]{revtex}
\begin{document}
\tightenlines
\title{Superconformal Quantum Mechanics via Wigner-Heisenberg
Algebra}
\author{H. L. Carrion$^{\ddagger}$ $^{\spadesuit}$ and R. de Lima
Rodrigues$^{\alpha}$  \\ ${}^{\ddagger}, \alpha$ {\small Centro
Brasileiro
de Pesquisas F\'\i sicas}\\ {\small Rua Dr. Xavier Sigaud, 150}\\
{\small CEP 22290-180, Rio de Janeiro-RJ, Brazil}\\
${}^{\spadesuit}$ {\small Instituto de F\'{\i}sica, Universidade
Federal do Rio de Janeiro } \\ {\small caixa postal 68528, 21945-970 Rio de Janeiro} \\
{\small RJ, Brazil}.\\ ${}^{\alpha}$ {\small Departamento de
Ci\^encias Exatas e da Natureza } \\ {\small Universidade Federal
da Para\'\i ba, Cajazeiras -- PB, 58.900-000, Brazil} }

\maketitle

\begin{abstract}
We show the natural relation between the Wigner Hamiltonian and
the conformal Hamiltonian. It is presented a model in
(super)conformal quantum mechanics with (super)conformal symmetry
in the Wigner-Heisenberg algebra picture $ [x,p_{x}]= i(1+c{\bf
P})$ (${\bf P}$ being the parity operator). In this context, the
energy spectrum, the Casimir operator, creation and annihilation
operators  are defined. This superconformal Hamiltonian is similar
to the super-Hamiltonian of the Calogero model and it is also an
extension of the super-Hamiltonian for the Dirac Oscillator.
\end{abstract}

\vspace{2cm} {\em E-mails:}{ hleny@cbpf.br; rafael@df.ufcg.edu.br;

\newpage
\pagestyle{plain} \renewcommand{\thefootnote}{\arabic{footnote}}

\section{Introduction}

\paragraph*{}

The problem of unifying quantum mechanics and gravity is one of
the greatest unsolved problems in physics. In this context, the
quantum mechanical black holes provide an arena in which quantum
mechanics and gravity meet head on. The conformally-invariant
quantum mechanical model was investigated firstly by Alfaro {\it
et al.} \cite{alfaro76} and an extended superconformal quantum
mechanics has recently been discovered as a superconformal
structure in multi black-hole quantum mechanics (see
\cite{stroming,gozzi00,AP,FR} and references therein). Also, some
super-conformal models are sigma models that describe the
propagation of a non-relativistic spinning particle in a curved
background \cite{RC}. It was recently conjectured by Gibbons and
Townsend that large $n$ limit of an $N=4$ superconformal extension
of the $n$ particle Calogero model \cite{GT} might provide a
description of the extreme Reissner-Nordstr\"{o}m black hole near
the horizon. In addition, the relation between the superconformal
mechanics and the nonlinear supersymmetry we can see in \cite{CM}.

In 1950, Wigner\cite{Wigner50} proposed the interesting question,
"Do the equations of motion determine the quantum-mechanical
commutation relations?" and he found as an answer a generalized
quantum rule for the one-dimensional harmonic oscillator. In the
next year, Yang \cite{Yang} found the coordinate representation
for the linear momentum operator which realizes this
aforementioned generalized quantum rule. Yang's wave mechanical
description was further studied by Ohnuki {\it et al.}
\cite{Ohnuki} and Mukunda {\it et al.} \cite{Sharma}.
Jayaraman-Rodrigues have identified the free parameter of the
Celka-Hussin's model with that of the Wigner parameter of a
related super-realized general 3D Wigner oscillator system
satisfying a generalized (super) quantum commutation relation of
the $\sigma_3$-deformed Heisenberg algebra \cite{JR90,JR94}.
Recently, the deformed Wigner-Heisenberg (WH) oscillator algebra
has been investigated in the context of the generalized statistics
(introduced in physics in the form of parastatistics as an
extension of the Bose and Fermi statistics) \cite{Mi97,Mi00}. On
the other hand, the elements of the conformal group can be
represented in terms of ladder operators of deformed quantum
oscillators  \cite{elr01} and the WH-algebra  has also been
investigated in connection with noncommutative geometry
\cite{palev03,palev03b}.

In this paper, firstly  we found the simple connection between the
Wigner Hamiltonian and the conformal Hamiltonian of  Ref.
\cite{stroming}.  We proceed by showing the interesting new
structures in conformal quantum mechanics in the WH picture. It is
introduced the new well defined conformal Hamiltonian (\ref{L0}),
its energy spectrum,  the Casimir operator, raising (or creation)
and the lowering (or annihilation) operators using the
Wigner-Heisenberg algebra. It is shown, for example, that the
eigenvalues of this conformal Hamiltonian is dependent of the
Wigner parameter  $c$ and the eigenvalues of the parity operator
$P$. When $c$=0 we obtain the usual conformal Hamiltonian
structure. Moreover,
 we present the new superconformal Hamiltonian with
Wigner-Heisenberg algebra structure. In the same way, we study the
energy spectrum and  construct the supersymmetric Casimir
operator. Our motivation is its potential application in (multi)
black-hole quantum mechanics \cite{stroming,GT} with the
possibilities to introduce new structures in this subject.

This work is organized as follows. In Sec. II we start by
summarizing the essential features of the formulation of ${\bf
P}$-deformed Wigner-Heisenberg oscillator algebra. We discuss the
connection between the usual conformal Hamiltonian and the Wigner
Hamiltonian. In Sec. III we present the conformal quantum
mechanics based on the  WH-algebra and discuss the eigenvector
problem by defining the Casimir operator, the ladder and the
annihilation operators. In Sec. IV we present the superconformal
Hamiltonian based  on the WH-algebra picture  and
 define the supercharged operator in the Yang representation. We
compute the superconformal algebra and discuss the eigenvalue and
the eigenvector problems in the supersymmetric context. As in the
bosonic case, in the supersymmetric case we defined the raising
and the lowering  operators and then we discuss the eigenvalue
problem for the well defined superconformal Hamiltonian. In
figures 1 and 2 the supersymmetric potentials for each specific
parity operator eigenvalue are plotted.

\section{The P-Deformed Heisenberg Algebra}

\paragraph*{}

In order to make the paper self-contained we present a short
discussion of the $P$-deformed Heisenberg algebra. The Wigner
Hamiltonian expressed in the symmetrized bilinear form in terms of
the mutually adjoint operators $\hat a{^\pm},$ is defined by

\begin{equation}
\label{w1} \hat H{_W} = \frac{1}{2}( \hat {p}^{2}_x+ \hat x^{2} )
= \frac{1}{2}[\hat a^{-}, \hat a^{+}]_+ = \frac{1}{2}(\hat a^{-}
\hat a^{+} + \hat a^{+} \hat a^{-}),
\end{equation}
where

\begin{equation}
\label{w2} \hat {a}^{\pm} = \frac{1}{\sqrt {2}} (\pm i\hat {p}_x -
\hat x).
\end{equation}
Wigner showed that the Heisenberg's equations of motion

\begin{equation}
\label{w3} [\hat {H}_{W}, \hat {a}^{\pm}]_- = \pm \dot{\hat
{a}}^{\pm},
\end{equation}
do not necessarily entail in the usual quantum rule

\begin{equation}
\label{RCC} [\hat{a}^{-}, \hat{a}^{+}]_{-} = 1 \Rightarrow [\hat
x,\hat p_{x}]_{-} =i, \quad \hbar=1,
\end{equation}
but a more general quantum rule \cite{Yang,Ohnuki,Sharma} given by

\begin{eqnarray}
\label{CW} [\hat a{^-}, \hat a{^+}]_- = 1 + c \hat R
\Longrightarrow [\hat {x}, \hat {p}_x]_- = i (1 + c \hat {R}),
\end{eqnarray}
where $c$ is a real constant, called Wigner parameter, related to
the ground state energy $E^{(0)}_W \geq 0$ by virtue of the
positive semi-definite form of $\hat H_W$\footnote{Note that the
case $c=0$ corresponds to the usual oscillator with
$E^{(0)}=\frac{1}{2}, \quad \hbar=\omega=1.$}

\begin{equation}
\label{w5} \vert c \vert = 2E^{(0)}_W - 1.
\end{equation}

The basic (anti-)commutation relations (\ref{w1}) and (\ref{w3}),
together with the derived relation (\ref{CW}) constitute ${\bf
P}$-deformed Wigner-Heisenberg algebra or in short  the
WH-algebra. The WH-algebra can also be obtained by the requirement
that $\hat x$ satisfies the classical equation of motion
($\ddot{\hat x} + \hat x = 0$).

Note that $\hat R$ is an abstract operator satisfying the
properties

\begin{equation}
\label{w6} [\hat R, \hat {a}^{\pm}]_+ = 0 \Rightarrow [\hat {R},
\hat{H}_{W}]_- = 0; \quad \hat R^{\dagger} = \hat {R}^{-1} = \hat
R, \quad \hat {R}^2 = 1.
\end{equation}
Also, we have

\begin{eqnarray}
\label{w7}
\hat{H}_W &=& \hat{a}^{+}\hat{a}^{-} + \frac{1}{2}(1 + c \hat{R})\nonumber\\
&=& \hat{a}^{-}\hat{a}^{+} - \frac{1}{2}(1 + c \hat{R}).
\end{eqnarray}
In the mechanical representation first investigated by Yang
\cite{Yang}, $\hat R$ is realized by the parity operator ${\bf
P}:$

\begin{equation}
\label{w8} {\bf P}|x> = |-x> \Rightarrow [ {\bf P}, x ]_{+} = 0 ,
\quad [ {\bf P} , p_x ]_{+} = 0 , \quad {\bf P}^{\dagger}={\bf
P}^{-1} = {\bf P}, \quad {\bf P}^2=1.
\end{equation}
Indeed, Yang \cite{Yang} found the coordinate representation for
the momentum operator $p_x$ as given by

\begin{eqnarray}
\label{w9}
\hat p{_x} \longrightarrow p{_x} = p +
i\frac{c}{2x}{\bf P}, \quad \hat x \longrightarrow x, \quad
p=-i\frac{d}{dx}
\end{eqnarray}

\begin{equation}
\hat a^{\pm} \longrightarrow a^{\pm}_{\frac c2}=
\frac{1}{\sqrt{2}} \left(\pm \frac{d}{dx}\mp\frac{c}{2x}{\bf P} -
x \right).
\end{equation}
Yang's wave mechanical description was further investigated in
\cite{Ohnuki,Sharma}.

The ${\bf P}$-deformed Heisenberg algebra is based on the
following general quantum rule \cite{Yang}

\begin{equation}  \label{newqr}
\label{PD} [\hat a{^-}, \hat a{^+}]_- = 1 + c {\bf P}
\Longrightarrow [\hat {x}, \hat {p}_x]_- = i (1 + c {\bf P}),
\end{equation}
where

\begin{equation}
\label{PR} [{\bf P}, \hat {a}^{\pm}]_+ = 0 \Rightarrow [{\bf P},
\hat{H}_{W}]_- = 0.
\end{equation}
When we replace the equation (\ref{w9}) into the equation
(\ref{w1}) for the Wigner Hamiltonian we obtain

\begin{eqnarray}
\label{osb} \hat{H}_{W-} = {1\over 2}\left\{ p^{2} + x^{2} +
{1\over 4x^2} (c^2+ 2c)\right\}
\end{eqnarray}
where the parity operator has been taken the value $-1$.

If we consider the following conformal Hamiltonian (equation (2.6)
of  ref. \cite{stroming})
\begin{eqnarray}
\label{L01} L_0= \frac{p^2}{2}+\frac{g}{2x^2}+\frac{x^2}{2},
\end{eqnarray}
and  choose $ c^2+ 2 c = 4 g $ in (\ref{osb}) then the Wigner
Hamiltonian is equal the conformal Hamiltonian (\ref{L01}) in the
coordinate representation

\begin{equation}
L_{0} = \hat{H}_{W_-},
\end{equation}
for the case ${\bf P}\rightarrow -1.$ This is a simple observation
but leads to new interesting results.  Note that introducing the
more general quantum rule (\ref{CW}) we are providing conformal
symmetry to the Wigner Hamiltonian. The energy spectrum of the
Hamiltonian $L_{0}$ (or $\hat{H}_{W-}$) is well defined. The
eigenstates of $L_{0}$ form an infinite tower above the {\it
ground state}, in integer steps \cite{alfaro76,stroming}.

On the other  hand,  it is important to comment that there exists
a hidden supersymmetry in the  WH-algebra structure. If we define
as the bosonic generators $\{\hat{H}_{W}, (\hat{a}^{+})^2 ,
(\hat{a}^{-})^2 \}$ and the fermionic ones as
$\{\hat{a}^{+},\hat{a}^{-} \}$ then they close the $ osp(1/2)$
superalgebra.

\section{Conformal symmetry in the Wigner-Heisenberg picture}

\paragraph*{}

In this section we study the Hamiltonian with conformal symmetry
in the WH-algebra picture. This is  a modified version of the
usual conformal Hamiltonian with a standard canonical quantum rule
$[\hat{x},\hat{p}_{x}]=\imath $ \cite{alfaro76,stroming}. Let us
define the new Hamiltonian

\begin{equation}
 \label{ham0} H = \frac{p_{x} ^{2}}{2}+ \frac{g}{2 x^2},
\end{equation}
where $g$ is a coupling constant and  $p_{x}$ is defined in
(\ref{w9}) with new quantum rule (\ref{newqr}). This Hamiltonian
(\ref{ham0}) commutes with the parity operator ${\bf P}.$

Next let us introduce the operators

\begin{eqnarray} \label{DK}
D&&=\frac{p_x x+xp_x}{2}, \nonumber\\
K&&= \frac{x^2}{2}.
\end{eqnarray}
We can demonstrate that these three operators satisfy the
following $SL$(2,{\bf R}) algebra

\begin{equation} \label{alg}
\label{scalg2} [H, D]= -2i H, \quad [H, K]= -i D, \quad [K, D]=
2iK,
\end{equation}
where $D$ is known to be the generator of dilatation (it generates
the re-scalings $x\rightarrow \gamma x $, $p_{x} \rightarrow
p_{x}/\gamma$) and $K$ is the generator of special conformal
transformations. Since $D$ and $K$ do not commute with the
Hamiltonian $H$, they do not generate symmetries in the usual
sense of relating the degenerate states. Rather they can be used
to relate states with different eigenvalues of $H$ (\ref{ham0}).

It is possible to show in any quantum mechanics with operators
obeying the $SL$(2,{\bf R}) algebra (\ref{alg}), that if
$|\chi\rangle$ is a state of energy $E$, then $e^{\imath \alpha
D}$ is a state for the energy $ e^{2 \alpha} E.$ Thus, if there is
a state of nonzero energy then the spectrum is continuous. Note
that this result provides the spectrum of Hamiltonian $H$
(\ref{ham0}) to be continuous and then its eigenstates are not
normalizable.

 Let us consider the following linear combinations

\begin{eqnarray} \label{LL0}
L_{+}&=& \frac 12(H-K+iD),\nonumber\\ L_{-}&=& \frac
12(H-K-iD),\nonumber\\ L_0&=& \frac 12(H+K).
\end{eqnarray}
The generators (\ref{LL0})  satisfy the $SL$(2,{\bf R}) algebra
(Virasoro form), with the following commutation relations

\begin{equation} \label{calg1}
[L_{-}, L_+]= 2L_0, \quad [L_{0}, L_+]= +L_{+} \quad [L_0, L_-]=
-L_-.
\end{equation}
With the definition (\ref{DK}), (\ref{alg}) and (\ref{LL0}), we
have

\begin{eqnarray}
\label{L0} L_0= \frac{p_x^2}{4}+\frac{g}{4x^2}+\frac{x^2}{4}.
\end{eqnarray}
The potential energy for this operator (\ref{L0}) have an absolute
minimum, then it has a discrete spectrum with normalized
eigenstates. If $L_0$ satisfies the following eigenvalue equation

\begin{equation}
L_0|n>= \epsilon_{n}|n>,
\end{equation}
and using the algebra (\ref{calg1}) we see that $L_{-}$ and
$L_{+}$ form the creation and annihilation operators for the
Hamiltonian $L_{0},$ so that

\begin{equation}
\label{en} \epsilon_{n}= \epsilon_0 +n, \quad n= 0, 1, 2,\cdots_.
\end{equation}

In  other words, the eigenvalues of $L_{0}$ form an infinite tower
above ``the ground state", in integer steps.

The $SL$(2,{\bf R}) Casimir operator is given by

\begin{equation}
L^2= L_{0}(L_0 - 1) - L_+L_- ,
\end{equation}
where

\begin{equation}
 [L^2, L_+]= [L^2, L_-]= [L^2, L_0]= 0.
\end{equation}

This Casimir operator in terms of the generators of the conformal
group becomes

\begin{equation}
L^2= \frac{HK + KH}{2}-\frac{D^2}{4}.
\end{equation}
Using the WH-algebra,  the action of this Casimir operator on the
eigenstates $|n>$ gives us

\begin{equation}
L^2= \frac{1}{16}[4g - 3 - c(2{\bf P}-c)]
\end{equation}
or

\begin{equation} \label{casim2}
L^2= \left\{
\begin{array}{c}
\frac{1}{16}[4g - 3 - 2c + c^2], \quad {\bf P}\rightarrow +1, \\
\frac{1}{16}[4g - 3 + 2c + c^2], \quad {\bf P}\rightarrow -1.
\end{array}\right.
\end{equation}
Defining $l_{0}$ by  $ L^2 = l_{0} (l_{0}-1)$,   from the relation
(\ref{casim2}) we obtain the ``ground state" for this Casimir
operator

\begin{equation}
\ell_0= \left\{
\begin{array}{c}
\frac{1\pm \sqrt{g + \frac {1}{4}(1 - c)^2}}{2}, \quad {\bf
P}\rightarrow 1\\ \frac{1\pm\sqrt{g + \frac {1}{4}(1+ c)^2}}{2},
\quad {\bf P}\rightarrow -1
\end{array}\right.
\end{equation}
and one can see that the coupling constant defined in (\ref{ham0})
must satisfy ( $g> -\frac{1}{4}(1 \pm c )^2  $).

When the constant $c$  vanishes the Casimir operator becomes $L^2=
\frac{1}{16}(4g - 3),$ so that the ground state has the eigenvalue
$\ell_0= \frac{1}{2}(1\pm\sqrt{g+\frac 14}),$ which is a well
known result for the usual conformal mechanics with the canonical
commutation relation $[x,p]=i$ \cite{stroming}.

Note that $\epsilon_{0} = \ell_0$, or   $\epsilon_{0} = 1- \ell_0
$ and

\begin{eqnarray}
L_+ |n,\ell_0>&=& \sqrt{(\epsilon_n + \ell_0)(\epsilon_n -\ell_0
+1)} |n+1, \ell_0>, \label{Lmas}  \nonumber \\ L_- |n,\ell_0>&=&
\sqrt{(\epsilon_n - \ell_0)(\epsilon_n +\ell_0-1)} |n-1, \ell_0>.
\label{Lmenos}
\end{eqnarray}

From Eqs.  (\ref{PD}) and  (\ref{LL0}), we can express $L_{+}$ and
$L_{-}$ in terms of the operators $\hat{a}_{+}$ and $\hat{a}_{-} $

\begin{eqnarray}
L_{+} &=& - \frac{1}{2} {\hat{a}_{+}}^2 + \frac{g}{4x^2}, \\ L_{-}
&=& - \frac{1}{2} {\hat{a}_{-}}^2 + \frac{g}{4x^2}.
\end{eqnarray}

\section{The superconformal quantum mechanics}

\paragraph*{}

The superconformal quantum mechanic has been examined in
\cite{stroming}-\cite{RC}. Conformal sigma models may have
applications in the context of $AdS/CFT$ correspondence with
$AdS_{2} \times M$ background \cite{BGIK}. Another application for
these models is in the study of the radial motion of test particle
near the horizon of extremal Reissner-Nordstr\"{o}m black holes
\cite{RC,GT}. Also, another interesting application of the
superconformal symmetry is the treatment  of the Dirac oscillator
\cite{moreno,MR}.

In this section we introduce the explicit supersymmetry  for the
conformal Hamiltonian in the WH-algebra picture. Consider the
supersymmetric generalization of $H$ (Eq. (\ref{ham0})) given by

\begin{equation}
{\cal H} = \frac{1}{2} \{Q_{c}, Q_{c}^{\dagger} \},
\end{equation}
where the new supercharge operators are given in terms of the
momentum Yang representation

\begin{eqnarray}
Q_{c} &&= \left(- i p_{x} + \frac{\sqrt{g}}{x}\right)
\Psi^{\dagger}, \nonumber\\ Q_{c}^{\dagger} &&= \Psi \left(i p_{x}
+
\frac{\sqrt{g}}{x}\right), \nonumber\\
\end{eqnarray}
with $\Psi$ and $\Psi^{\dagger}$ being Grassmannian operators so
that its anticommutator is $\{\Psi, \Psi^{\dagger}\}=
\Psi\Psi^{\dagger} + \Psi^{\dagger} \Psi= 1.$

Explicitly the superconformal Hamiltonian becomes
\begin{equation}
{\mathcal H} = \frac{1}{2} ( {\bf 1}p_{x}^2 + \frac{{\bf 1}g+
\sqrt{g}B(1-c {\bf P})}{x^2}) \label{Hsussy}
\end{equation}}
where

\begin{equation}
B =\left[\Psi^\dagger, \Psi \right],
\end{equation}
so that the parity operator is conserved, i.e.,
$\left[\mathcal{H}, {\bf P} \right] = 0.$

When one introduces the following operators
\begin{eqnarray}
S &=& x \Psi^\dagger, \nonumber\\
S^\dagger &=& \Psi x,
\end{eqnarray}
it can be shown that these operators together with the conformal
quantum mechanics operators $D$ and $K$

\begin{eqnarray}
D &=& \frac{1}{2} (x p_{x} + p_{x} x ), \nonumber\\
K &=& \frac{1}{2} x^2,
\end{eqnarray}
satisfy the deformed superalgebra $osp(2|2)$\footnote{Actually,
this superalgebra is $osp(2|2)$ when we fix ${\bf P}=1$ or $ {\bf
P}=-1$.}, viz.,

\begin{eqnarray}
\left[ {\mathcal{H}}, D \right] &=& -2\imath \mathcal{H} , \nonumber\\
\left[{\mathcal{H}},K \right] &=& -\imath D ,\nonumber\\
\left[K, D \right] &=& 2 \imath K,\nonumber\\
\{Q_{c}, Q_{c}^\dagger \} &=& 2 \mathcal{H},\nonumber \\
\{Q_{c}, S^\dagger\} &=& -\imath D -
\frac{1}{2} B(1+ c {\bf P}) + \sqrt{g}, \nonumber\\
\{Q_{c}^\dagger, S \} &=& + \imath D - \frac{1}{2} B(1+ c {\bf P})
+
\sqrt{g}, \nonumber\\
\left[Q_{c}^{\dagger}, D \right] &=& -\imath Q_{c}^\dagger,\nonumber\\
\left[Q_{c}^\dagger, K \right] &=& S^\dagger, \nonumber\\
\left[Q_{c}^\dagger, B \right] &=& 2 Q_{c}^\dagger, \nonumber\\
\left[Q_{c}, K \right] &=& -S,\nonumber\\
\left[Q_{c}, B \right] &=& -2 Q_{c}, \nonumber\\
\left[Q_{c}, D \right] &=& -\imath Q_{c}, \nonumber\\
\left[ \mathcal{H}, S \right] &=& Q_{c}, \hspace{2cm}
\left[ \mathcal{H}, S^\dagger \right] =-Q_{c}^\dagger, \nonumber\\
\left[ B, S^\dagger \right] &=& -2 S^\dagger,
\hspace{2cm} \left[B, S \right] = 2S, \nonumber\\
\left[D, S \right] &=& -\imath S, \hspace{2cm} \left[D, S^\dagger
\right] =
-\imath S^\dagger,\nonumber\\
\{S^\dagger, S \} &=& 2 K,
\end{eqnarray}
where, $\mathcal{H}, D, K, B $ are bosonic operators and $Q_{c},
Q_{c}^\dagger, S, S^\dagger$ are fermionic operators. This
superalgebra was introduced by Plyushchay in the bosonization of
supersymmetry context \cite{P}. The supersymmetric extension of
the conformal Hamiltonian $L_{0}$ (presented in  the previous
section) is

\begin{equation} \label{h0}
{\mathcal{H}}_{0} = \frac{1}{2} ( {\mathcal H} + K ), \quad
\left[{\mathcal{H}}_{0}, {\bf P} \right] = 0.
\end{equation}
The super-Hamiltonian $ {\mathcal{H}}_{0}$ is similar to the
supersymmetric Calogero interaction Hamiltonian associated with
two-particle interaction. The possibility that the $n$-particle
Calogero model \footnote{Calogero model \cite{Cal} describes the
system of $n$ bosonic particles   interacting through the inverse
square and harmonic potential, it is completely integrable (at
classical and quantum levels).} might be relevant for a
microscopic description of the extreme Reissner-Nordstrom black
hole, has been discussed in Refs. \cite{GT,AG}. Specifically, it
was conjectured by Gibbons and Townsend that the large $n$ limit
of $N=4$ superconformal version of the $n$-particle Calogero model
might provide microscopical description of the Reissner-Nordstrom
black hole near to the horizon \cite{GT}.
Continuing with our discussion, now we are able to build up the
ladder operators for the spectral resolution of the
super-Hamiltonian $ {\mathcal{H}}_{0}$.  Note that defining

\begin{eqnarray}
M &=& Q_{c}- S ,\\ Q &=& Q_{c} + S,
\end{eqnarray}
\begin{eqnarray}
 h &\equiv& \frac{1}{2} \{M, M^\dagger \} = 2
 {\mathcal{H}}_{0} + \frac{1}{2}(1+ c {\bf P} )B - \sqrt{g}, \nonumber\\
 \tilde{h}&\equiv& \frac{1}{2} \{Q, Q^\dagger \} = 2
 {\mathcal{H}}_{0} - \frac{1}{2}(1+ c {\bf P} )B + \sqrt{g},
\end{eqnarray}
we obtain

\begin{eqnarray}
 \left[ h, Q^\dagger \right ] &=& -2 Q^\dagger, \nonumber\\
 \left[ h, Q \right ] &=& 2 Q,
\end{eqnarray}
and
\begin{eqnarray}
 \left[ \tilde{h}, M^\dagger \right ] &=& 2 M^\dagger, \nonumber\\
 \left[ \tilde{h}, M \right ] &=& -2 M .
\end{eqnarray}
Thus we see that $Q$ and $Q^\dagger, \; M$ and $M^\dagger$ are
ladder operators associated to the SUSY Hamiltonian operators $h$
and $\tilde{h}$, respectively. We remark that since $h, \tilde{h}$
and ${\mathcal{H}}_{0}$ commute with each other, they form a set
of mutually commuting  operators. The superhamiltonian $h$ and
$\tilde{h}$ are extensions of the ones in references \cite{moreno,MR}
for the superconformal Dirac oscillator.

The supersymmetric generalizations of the ladder operators defined
in (\ref{LL0}) in the bosonic case, are given by
\begin{eqnarray}
 {\mathcal{L}}_{-}& \equiv & -\frac{1}{4} \{M, Q^\dagger \}
= -\frac{1}{2} (H-K - \imath D ), \nonumber\\
 {\mathcal{L}}_{+} &\equiv &  -\frac{1}{4} \{M^\dagger, Q \} =
 -\frac{1}{2} (H-K + \imath D )
\end{eqnarray}
with the following commutation relations of the $so(2,1)$ algebra

\begin{eqnarray}
 \left[{\mathcal{L}_{+}}, {\mathcal{H}}_{0} \right] &=& -\mathcal{L}_{+},
\nonumber\\
 \left[{\mathcal{L}}_{-}, {\mathcal{H}}_{0} \right] &=& + \mathcal{L}_{-},
\nonumber\\
 \left[\mathcal{L}_{+}, \mathcal{L}_{-} \right] &=& - 2{\mathcal{H}}_{0}.
\end{eqnarray}

The supersymmetric Casimir operator is given by

\begin{eqnarray} \label{CL2}
{\mathcal{L}}^2 =   {\mathcal{H}}_{0} ( {\mathcal{H}}_{0}-1 ) -
\mathcal{L}_{+} \mathcal{L}_{-}.
\end{eqnarray}

Thus, the energy spectrum becomes
\begin{eqnarray}
{\mathcal{H}}_{0} \mid m,s > &=& \zeta_{m} \mid m,s>,
  \label{Eau} \nonumber\\
{\mathcal{L}}^2 \mid m,s > & = & {s} (s-1)\, \mid m,s >,
\end{eqnarray}
with $ \zeta_{m} = \zeta_{0} + m $.

 Since the eigenket $\mid m,s >$ is normalized, from
(\ref{CL2}) and (\ref{Eau}) we obtain $s = s_{0} = \zeta_{0}.$ In
this case, we may write the analogous representations to the
ladder operators given by Eq. (\ref{Lmas})

\begin{eqnarray}
{\mathcal{L}}_{+}  \mid m,s > &=& \alpha_{m+1}  \mid m+1,s >, \nonumber\\
{\mathcal{L}}_{-}  \mid m,s > &=& \beta_{m-1}  \mid m-1,s>
\end{eqnarray}
where $\alpha_{m+1}$, and $ \beta_{m-1}$ have a similar form to
the bosonic case, given in the equations (\ref{Lmas}).

Explicitly

\begin{eqnarray}
{\cal L}^2 = \frac{1}{16} [( 4 g -3 -c (2{\bf
P}-c)){\mathbf{l}}_{2\times 2} + 4 \sqrt{g}B(1-c{\bf P}) ]
\end{eqnarray}
where

\begin{eqnarray}
{\mathbf{l}}_{2\times 2} &=&  \left[
\begin{array}{cc}
  1 & 0\\
  0 & 1
\end{array} \right],  \hspace{3cm} B = \left[
\begin{array}{cc}
  -1 & 0\\
  0 & 1
\end{array} \right]= -\sigma_{3}, \label{const}
\end{eqnarray}
and the parity operator must be substituted by its eigenvalues on
the eigenstates $\mid m,s>.$ In the case  ${\bf P} \rightarrow
+1,$ on gets

\begin{equation}
{\cal L}^2 = \frac{1}{16}\left[
\begin{array}{cc}
  4 g -3 -2c + c^2 -4 \sqrt{g}(1-c ) & 0\\
  0 &  4 g -3 -2c + c^2 +4 \sqrt{g}(1-c )
\end{array} \right],
\end{equation}
with the following eigenvalues

\begin{eqnarray}
s(s-1) &=& \frac{1}{16} (4 g -3 -2c + c^2 -4 \sqrt{g}(1-c )) \\
s^{\prime}(s^{\prime}-1) &=& \frac{1}{16} (4 g -3 -2c + c^2 +4
\sqrt{g}(1-c )).
\end{eqnarray}
 From equation  (\ref{h0})  and (\ref{const})  we
obtain the SUSY conformal Hamiltonian in the coordinate
representation

\begin{eqnarray} \label{hs01}
{\mathcal{H}}_{01} = \frac{1}{4}\left[
\begin{array}{cc}
  p^2 + x^2 + \frac{c^2+2c+4g -4\sqrt{g}(1-c)}{ 4x^2}  & 0\\
  0 &  p^2 + x^2 + \frac{c^2+ 2c +4g + 4g \sqrt{g}(1-c)}{4 x^2}
\end{array} \right],
\end{eqnarray}
with the following superconformal potential

\begin{eqnarray} \label{v1}
 V_{01}(x) & \equiv& \left[
\begin{array}{cc}
  V_{01} (+)  & 0\\
  0 &   V_{01} (-)
\end{array} \right] =  \frac 14\left((x^2+\frac{c^2+2c+4g}{4x^2}){\bf 1}_{2\hbox{x}2} -
 \sigma_3\frac{\sqrt{g}(1-c)}{ x^2}\right).
\end{eqnarray}
In Fig. 1 we plot  $ V_{01} (+)$, the bosonic part of this
supersymmetric potential.

In the case ${\bf P} \rightarrow -1$ we get

\begin{eqnarray}
{\cal L}^2 = \frac{1}{16}\left[
\begin{array}{cc}
  4 g -3 +2c + c^2 -4 \sqrt{g}(1+c ) & 0\\
  0 &  4 g -3 +2c + c^2 +4 \sqrt{g}(1+c)
\end{array} \right].
\end{eqnarray}

Also, in this case

\begin{eqnarray} \label{hs02}
{\cal H}_{02} = \frac{1}{4}\left[
\begin{array}{cc}
 p^2 + x^2 + \frac{c^2 -2c+4g-4\sqrt{g}(1+c)}{4 x^2}& 0\\
  0 &  p^2 + x^2 + \frac{c^2-2c+4g+4\sqrt{g}(1+c)}{4 x^2}
\end{array} \right]
\end{eqnarray}
with superconformal potential

\begin{eqnarray}\label{v2}
 V_{02}(x) &=&  \frac 14 \left( (x^2+ \frac{c^2-2c+4g}{4x^2}){\bf 1}_{2\hbox{x}2}-
\sigma_3\frac{\sqrt{g}(1+c)}{ x^2} \right).
\end{eqnarray}
In Fig. 2 we plot $V_{02} (+)$, the bosonic  part of this
supersymmetric potential.

Observe that we are using the convention that the operator number
$N_f$  has the fermion number $n_f=0$ and is associated to the
bosonic state and this eigenstate is given by
\begin{eqnarray}
\left(
\begin{array}{c}
  1 \\
  0
\end{array}
\right).
\end{eqnarray}
 Similarly, the eigenstates of $N_{f}$ with the fermion number
$n_f=1$ is associated to the fermionic state, and is given by
\begin{eqnarray}
\left(
\begin{array}{c}
  0 \\
  1
\end{array}
\right).
\end{eqnarray}

Finally, observing the supersymmetric Hamiltonians ${\cal H}_{01}$
and $ {\cal H}_{02}$ we recall that the parameter $c$ is real and
$g$
 must be positive. With appropriate choice of these parameters we
can recover various cases,  for example the oscillator type
superhamiltonian (without the presence of the proportional term
$1/x^2$) although the bosonic part without supersymmetry has this
term  $(g \neq 0 )$. In this case the supersymmetric potential
will not have a singularity at the origin of coordinates.

\section{Conclusion}

\paragraph*{}

In this work, we have analyzed  the connection between the
conformal quantum mechanics \cite{alfaro76,stroming} and the
Wigner-Heisenberg algebra \cite{Yang,Ohnuki,JR90}. With an
appropriate relationship between the coupling constant $g$ and
real constant $c$ one can identify the Wigner Hamiltonian  with
the conformal Hamiltonian $L_{0}$. The important result is that
the introduction of the Wigner-Heisenberg algebra in the conformal
quantum mechanics is still consistent with the  conformal
symmetry. The spectrum for the Casimir operator, the Hamiltonian
$L_{0}$ and the ladder operators depend on the parity operator. We
also investigated the supersymmetrization of this model, in that
case we obtain a new spectra for the supersymmetric and conformal
Hamiltonian of the Calogero interaction's type. In this case the
spectrum for the super-Casimir operator and the superhamiltonian
depend also on the parity operator. One motivation for this work
is the future applications in the problems related to black holes
\cite{stroming,GT} and the construction of a supersymmetric
quantum mechanics with conformal symmetry for the n-particles in
the Wigner-Heisenberg picture. Let us point out that one can
consider the same analysis implemented in this work for the Dirac
oscillator, getting a generalization of the works
\cite{moreno,moshi,rv03do}.

Finally, it would be interesting to investigate the possible
connection  between the hidden supersymmetry in the WH-algebra
structure  and the  explicit supersymmetry implemented in the
section 4, work in  this direction is under current research.

\vspace{1cm}

\centerline{\bf ACKNOWLEDGMENTS}

\vspace{0.5cm}

The authors are grateful to Prof. J. A. Helayel Neto for the kind
hospitality at CBPF-MCT. The authors  are also partially supported
by CNPq-CLAF. We are also  grateful to Harold Blas and F. Toppan
for interesting discussions and suggestions.



\vspace{0.0cm}
\begin{center}
\begin{figure}
\hspace{4.01cm}\scalebox{0.35}{\includegraphics*{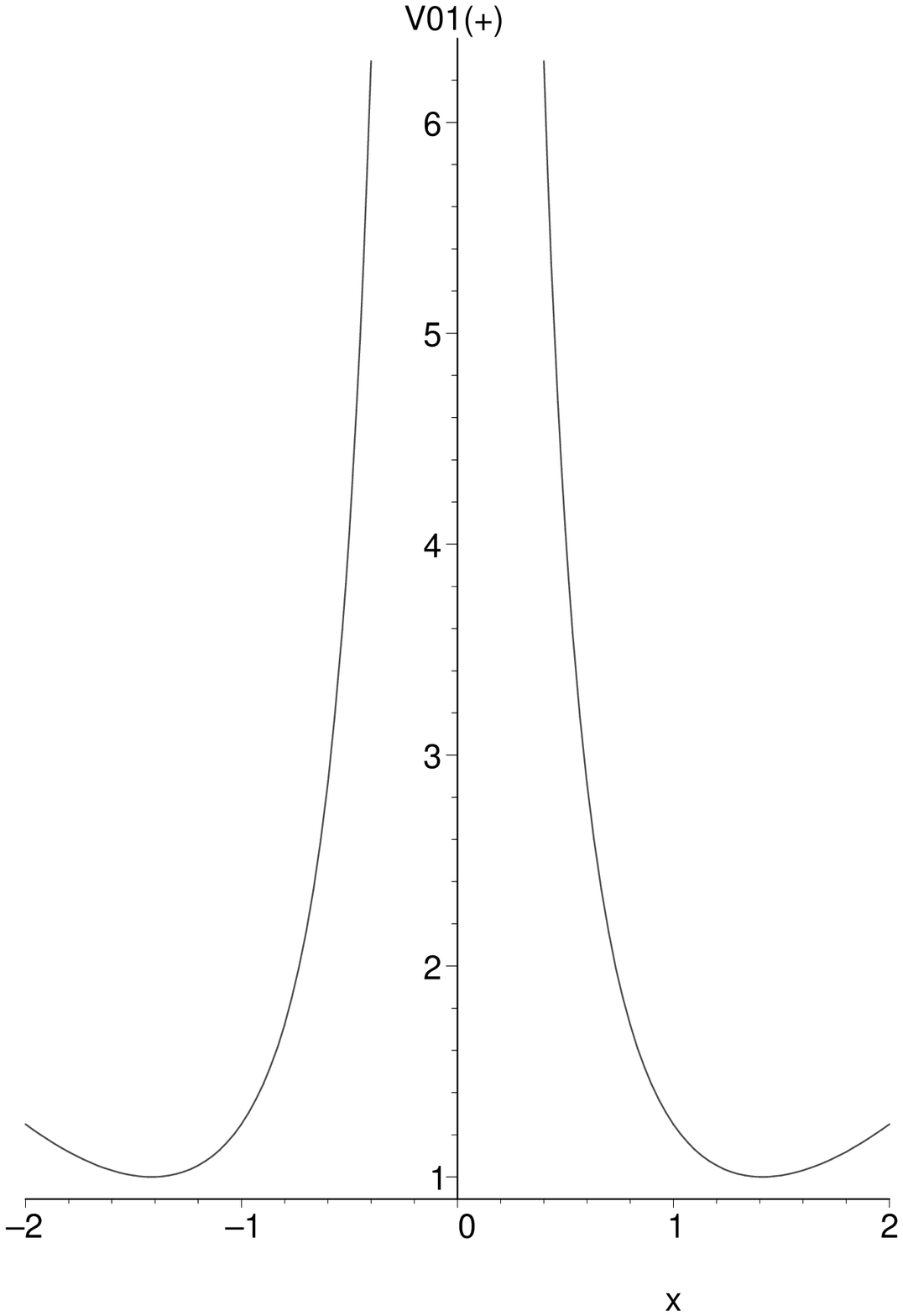}}\\ Fig
1. The bosonic part of the supersymmetric potential $V_{01},
\;$for$\;P=+1, c=2, g=1.$
\end{figure}
\end{center} \label{fig1}
\vspace{0.8cm}

\begin{center}
\begin{figure}
\hspace{4.01cm}\scalebox{0.35}{\includegraphics*{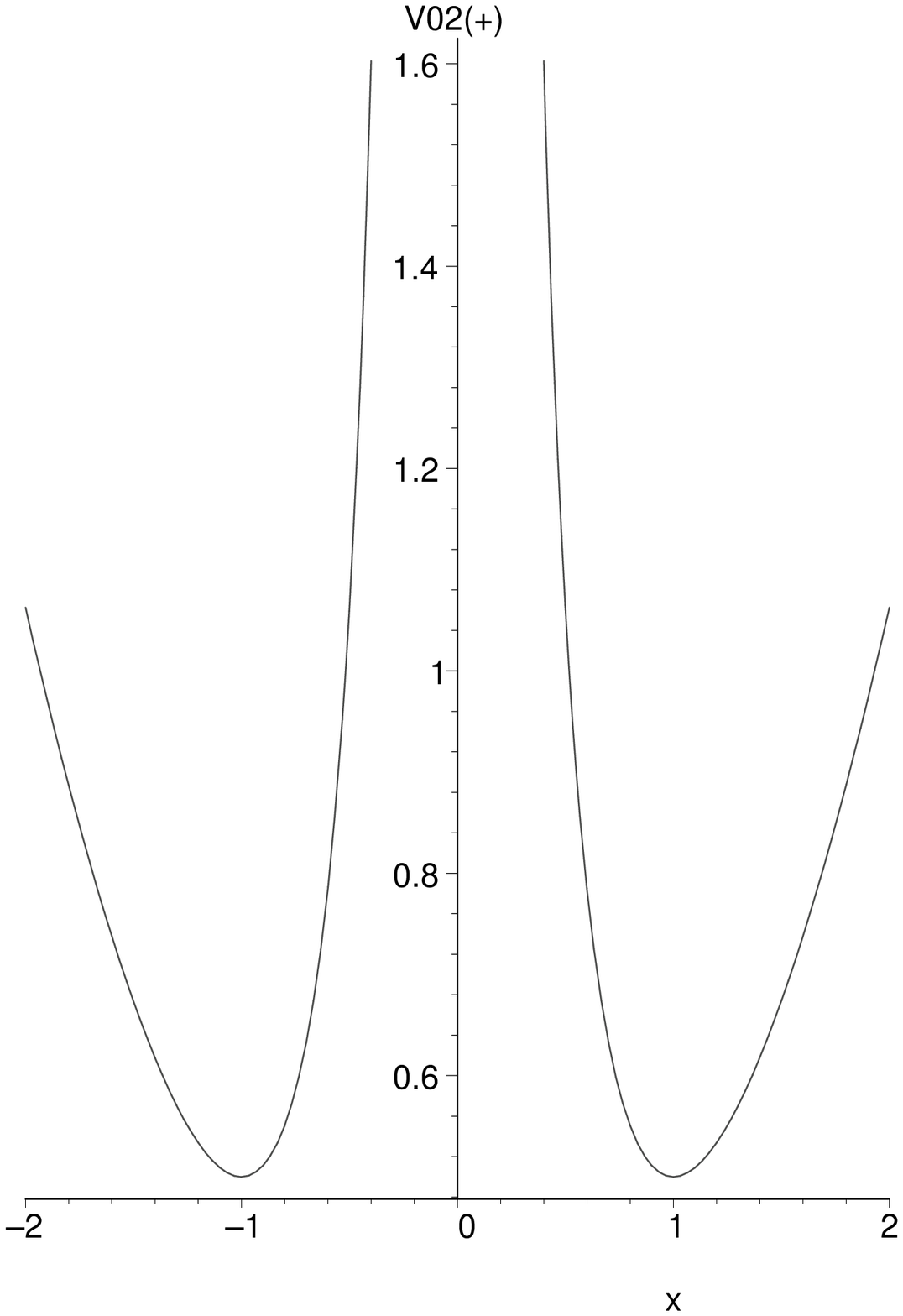}} \\Fig
2. The bosonic part of the supersymmetric potential $V_{02},\;$for
$\;P=-1, c=2, g=1.$
\end{figure}
\end{center} \label{fig2}


\begin{thebibliography}{99}

\bibitem{alfaro76} V. de Alfaro, S. Fubini and G. Furlan,
 {\it Nuovo Cimento} {\bf 34A} (1976), 569.

\bibitem{stroming} R. Britto-Pacumio, J. Michelson, A. Strominger
and A. Volovich, {\it Phys. Rev.} {\bf D50} (2000), 43,
het-th/9911066; hep-th/9908044.

\bibitem{gozzi00} E. Deodato, G. Furlan and E. Gozzi,
 {\it J. Math. Phys.} {\bf 41} (2000), 8083, hep-th/9910220.

\bibitem{AP} V. P. Akulov and I. A. Pashnev, {\it Theor. Math. Phys.} {\bf
56} (1983), 862.

\bibitem{FR} S. Fubini and E. Rabinovici, {\it Nucl. Phys.} {\bf  B
245} (1984), 17.

\bibitem{RC} P. Claus, M. Derix, R. kallosh, J. Kumar, P.
Townsend and A. Van Proeyen, {\it Phys. Rev. Lett.} {\bf 81}
(1998), 4553, hep-th/9804177.

\bibitem{GT} G. W. Gibbons and P. K. Townsend, {\it Phys. Lett.} {\bf
B454} (1999), 187.

\bibitem{CM} C. Leiva and M. S. Plyushchay, {\it  JHEP} {\bf
69} (2003), 0310.

\bibitem{Wigner50} E. P. Wigner, {\it Phys. Rev.} {\bf 77} (1950),
711.

\bibitem{Yang} L. M. Yang,
{\it Phys. Rev.} {\bf 84} (1951), 788.

\bibitem{Ohnuki} Y. Ohnuki and S. Kamefuchi, {\it J. Math. Phys.}
{\bf 19} (1978), 67; Y. Ohnuki and S. Watanabe,  {\it J. Math.
Phys.} {\bf 33} (1992), 3653.

\bibitem{Sharma} N. Mukunda, E. C. G. Sudarshan,
J. K. Sharma and C. L. Mehta, {\it J. Math. Phys.} {\bf 21}
(1980), 2386.

\bibitem{JR90} J. Jayaraman and R. de Lima Rodrigues,
 {\it J. Phys. A: Math. Gen.} {\bf 23} (1990), 3123.

\bibitem{JR94} J. Jayaraman and R. de Lima Rodrigues,
{\it Mod. Phys. Lett.} {\bf A9} (1994), 1047.

\bibitem{Mi97} M. S. Plyushchay,  {\it Nucl. Phys.} {\bf B491} (1997), 619;
M. S. Plyushchay, {\it Int. J. of Mod. Phys.} {\bf A15} (2000),
3679.

\bibitem{Mi00} M. S. Plyushchay,  {\it Deformed Heisenberg algebra with
reflection, anyons and supersimmetry of parabosons},
hep-th/0006238.

\bibitem{elr01} E. L. Da Gra\c{c}a, H. L. Carrion and R. de Lima Rodrigues,
 {\it Braz. J. Phys.} {\bf 33} (2003), 333, hep-th/0205167.

\bibitem{palev03} R. C. King, T. D. Palev and N. I. Stoilova and
J. Van der Jeugt, {\it J. Phys. A: Math. Gen.} {\bf 36} (2003),
4337, hep-th/0304136.

\bibitem{palev03b} R. C. King, T. D. Palev and N. I. Stoilova and
J. Van der Jeugt, {\it J. Phys. A: Math. Gen.} {\bf A36} (2003),
11999.

\bibitem{BGIK} S. Bellucci, A. Galajinsky, E. Ivanov and S.
Krivonos, {\it Phys. Lett.} {\bf B555} (2003), 99, hep-th/0212204.

\bibitem{moreno} R. P. Mart\'\i nez y Romero,
Mat\'\i as Moreno and A. Zentella, {\it Phys. Rev.} {\bf D43}
(1991), 2036.

\bibitem{MR} R. P. Martinez y Romero and A. L. Salas Brito, {\it J. Math.
Phys.} {\bf 33} (1992), 1831.

\bibitem{P} M. S. Plyushchay, {\it Mod. Phys. Lett.} {\bf A11}
(1996), 397.

\bibitem{Cal} F. Calogero {\it J. Math. Phys.} {\bf 10} (1969), 2197.

\bibitem{AG} A. V. Galajinsky, {\it Comments on N=4 Superconformal
Extension of the Calogero Model}, hep-th/0302156.

\bibitem{moshi} M. Moshinsky and A. Szczepaniac, {\it J. Phys. A: Math.
Gen.} {\bf 22} (1989), L817.

\bibitem{rv03do} R. de Lima Rodrigues and A. N. Vaidya,
{\it Dirac oscillator via R-deformed Heisenberg algebra},
Proceedings of the XXIII Brazilian National Meeting on Particles
and Fields (October/2002), site
www.sbfisica.org.br/eventos/enfpc/xxiii/procs/trb0013.pdf,
preprint{CBPF-NF-030/02}, hep-th/0301199.

\end{thebibliography}
\end{document}